\documentstyle[epsf,aasms4]{article}
\received{8 May 1998}
\accepted{25 August 1998}
\slugcomment{To appear in the Astrophysical Journal}
\begin{document}
\title{Infrared spectroscopy of NGC\,4151:\\
       ISO observations and NLR line profiles
      }

\author{E. Sturm\altaffilmark{1}, T. Alexander\altaffilmark{1,2,3}, 
D. Lutz\altaffilmark{1}, A. Sternberg\altaffilmark{2}, 
H. Netzer\altaffilmark{2}, and R.Genzel\altaffilmark{1}}

\altaffiltext{1}{Max-Planck-Institut f\"ur Extraterrestrische Physik, 
                 Postfach 1603, D-85740 Garching, Germany}
\altaffiltext{2}{Tel Aviv University, Ramat Aviv, Tel Aviv 69978, Israel}
\altaffiltext{3}{Institute for Advanced Study, Olden Lane, Princeton, NJ 08540,
                 USA}
\begin{abstract}
We present ISO-SWS and ISOPHOT-S 
spectroscopy\footnote{Based on observations 
       made with ISO, an ESA project with instruments funded by ESA member 
       states (especially the PI countries: France, Germany, The Netherlands 
       and the United Kingdom) and with the participation of ISAS and NASA.
}
of the
Seyfert galaxy NGC 4151. We detect a total of 17 fine-structure emission
lines emitted by a wide range of low- and high-excitation ions, two rotational 
lines of molecular hydrogen, and the Br$\beta$ HI line.

We find that the mid-IR fine-structure line profiles display blue asymmetries which
are very similar to those observed in the optical lines produced in the
narrow line region.
Because the mid-infrared
lines are much less sensitive to extinction than are the
optical lines this similarity places strong
constraints on scenarios which have been invoked to explain the optical line 
asymmetries.
For example, we are able
to rule out the simplest radial-motion-plus-dust scenarios for the production of
the line asymmetries.
Our preferred model is that of a central,
geometrically thin but optically thick, obscuring screen
of 
sub-arcsecond extension, enclosing a total hydrogen gas mass of $\gtrsim 
5\times10^6$
M${\odot}$. This mass may be molecular.

The weakness of `PAH' emission features in the low resolution
spectrum is evidence that star formation plays a minor role in the
circumnuclear region of NGC\,4151.
In a companion paper, we use the rich set of mid-infrared lines to determine the
obscured photoionizing continuum produced by the active galactic nucleus.

\keywords{galaxies: individual: NGC\,4151 --  galaxies: Seyfert -- 
line: profiles -- infrared: galaxies}
\end{abstract}
\section{Introduction}

NGC 4151 is one of the nearest (13.2 Mpc for $H_0=75$ km s$^{-1}$ Mpc$^{-1}$, 
1$^{\prime\prime}$=64 pc) 
and best studied Seyfert galaxies (Seyfert 1943; Oke \& Sargent 1968).
The optical spectrum is dominated by the central non-stellar
continuum source and the surrounding broad and narrow  emission line regions 
(BLR and NLR).
NGC 4151 is usually classified as a Seyfert 1 galaxy, though 
Osterbrock \& Koski (1976) classified
it as an intermediate Seyfert 1.5 system.
The central source is highly variable and the broad emission lines
reverberate rapidly (Penston \& P\'{e}rez 1984; Ulrich et al. 1991;
Edelson et al. 1996; Kaspi et al. 1996; Fernandez et al. 1997).  NLR emission is
produced within the inner $\sim 5^{\prime\prime}$ in a wide ionization cone
(Heckman \& Balick 1983; Boksenberg et al. 1995; 
Kaspi et al. 1996; Winge et al. 1997; Hutchings et al. 1998). The NLR is 
oriented along the direction of a non-thermal radio jet emanating from the 
nucleus (Pedlar et al. 1993). 
More extended ($\sim 30^{\prime\prime}$) narrow-line emission (EELR) is
produced in portions of the galactic disk which are
intersected by the ionization cone (Pedlar et al. 1993; Boksenberg et al. 1995).
High resolution optical imaging and spectroscopic studies show that the NLR
cloud dynamics are consistent with 
high-velocity outflows superposed on larger scale
galactic disk rotation (Schulz 1990; Veilleux 1991b; Boksenberg et al. 1995; 
Weymann et al. 1997; Winge et al. 1997; Hutchings et al. 1998).

NGC 4151 is one of many Seyfert galaxies whose optical NLR emission lines
display pronounced blue asymmetries and blue-shifts relative to the
systemic velocities (e.g. Heckman et al. 1981; Veilleux 1991a,c).
Such line asymmetries have generally been understood as being due to
a combination of gas inflow or outflow, coupled with selective dust extinction
of the red-shifted components.  Mid-infrared spectroscopic observations can
provide strong constraints on such models, since the mid-IR emission lines are
much less sensitive to extinction than are optical lines. 

In this paper we present 2.4-45 $\mu$m mid-IR spectroscopic observations
of NGC 4151 carried out with the {\it Infrared Space Observatory (ISO)}. 
We 
focus on
the remarkable similarity that we find between the shapes of the mid-IR and 
optical NLR line profiles.  In a companion paper (Alexander et al. 1998: A98)
we use our fine-structure emission line spectroscopy to determine the
shape of the (obscured) continuum radiation which photoionizes the NLR in 
NGC 4151. Our observations are part of our program to study the
mid-IR spectral properties of active galactic nuclei (AGNs),
starburst galaxies, and ultraluminous infrared galaxies (Lutz et al. 1996, 
1997; Kunze et al. 1996; Moorwood et al. 1996; Rigopoulou et al. 1996; 
Sturm et al. 1996; Genzel et al. 1998).

This paper is organized as follows.  In \S 2 we describe our observations.
In \S 3 we compare our mid-IR line profile observations with
optical line profiles, and show that they are very similar. 
In \S 4 we use this result to argue that the line asymmetries are likely
produced by a central and very optically thick obscuring screen, rather
than by dust which is distributed throughout the NLR or in the emitting
clouds themselves.  
We present a brief summary in \S 5.


\section{Observations}
\label{s:SWS}

We observed the central region of NGC\,4151 with the Short Wavelength
Spectrometer (SWS, de Graauw et
al. \cite{deGraauw}), and the spectrophotometer (ISOPHOT-S)
(Lemke et al. \cite{Lemke}) on-board the {\it ISO} (Kessler et al. 1996).
We also obtained a large aperture optical CCD spectrum using the Coud\'{e} 
Echelle spectrometer at the 2m telescope of the Karl-Schwarzschild-Observatory 
(KSO, Landessternwarte Th\"uringen). We use the optical observations
in our comparison analysis described in \S 3.
 
Our SWS observations were carried out during revolution 172 (1996 May 7).
We obtained a full grating
scan (SWS01 speed 1 mode -
$\lambda/\Delta\lambda \approx$ 250...600) from 2.4 to 45 $\mu$m as well as a
large set of individual line scans (SWS02 mode - $\lambda/\Delta\lambda
\approx$ 1000...2500). 
The full grating scan is relatively noisy, and we do not present it here.
The average total integration time per individual line scan was 
1200 seconds. We reduced our spectra using the SWS Interactive
Analysis System (SIA) (Lahuis et al. \cite{Lahuis}, Wieprecht et al. 
\cite{Wieprecht}, ). We adopted the September 1997 flux calibration. 

Our individual line spectra are shown in Figure
\ref{f:SWS-spectrum}. 
We detected 17 fine-structure emission lines from a wide range
of ionic species, two pure rotational
lines of molecular hydrogen (H$_2$) and the Br$\beta$ HI recombination line.
Table \ref{t:SWS-linefluxes} lists the measured line fluxes, as well as
upper limits for several undetected lines. 
The error estimates on observed fluxes in Table \ref{t:SWS-linefluxes} are
based on the uncertainty in defining the underlying continuum. In addition 
there is a general flux calibration uncertainty of $\approx$20\% (Schaeidt et 
al. \cite{Schaeidt}).
The SWS aperture sizes range between $14\arcsec\times20\arcsec$ and
$20\arcsec\times33\arcsec$.          
Yoshida and Ohtani (1993) showed that more than 80\% of the
[OIII] and [OII] optical line flux of the NLR plus EELR is produced within the 
inner
$\sim$4"x4" region. Therefore, we can safely assume that 
we have detected all of the mid-IR emission from the NLR and most of the EELR.
This is illustrated in Figure 
\ref{f:aperture},
in which the SWS apertures are overlayed on the [O III]
$\lambda$5007
contours of the NGC\,4151 nucleus (adapted from Yoshida \& Ohtani 1993).
We do not apply aperture 
correction factors to the measured line fluxes in our analysis.

As shown in Table 1
the observed fine-structure lines are emitted by ions with excitation energies
which range from
8 to 303 eV. The high-excitation coronal lines are 
unambiguous diagnostics of non-stellar photoionization (Oliva et al. 1994, 
Ferguson et al. 1997).
In A98 we use the fluxes listed in Table 1
together with
photoionization models to determine the spectral shape of the Lyman continuum
radiation field which is photoionizing the NLR.
As discussed in A98 the density sensitive flux ratios of the
[NeV] 14.32, 24.32 $\mu$m, [NeIII] 15.55,36.04 $\mu$m, and [SIII] 18.71,33.48
line pairs, in combination with the [O III] 51.81, 88.35 $\mu$m pair (Spinoglio
et al. 1998), imply an NLR electron density of $\sim 1000$ cm$^{-3}$.
The optical to mid-IR flux ratio [NeIII] 3868\AA/15.55$\mu$m implies an
electron temperature of $13000\pm2500$ K for the NLR gas, assuming a reddening
of E$_{B-V}$ = 0.05-0.15 (see A98).
The continua in our spectra agree well with the IRAS fluxes, but do not indicate
a pronounced bimodal dust emission as proposed by Rodriguez-Espinosa et al.
(1996). 

Our ISOPHOT-S observations were carried out during revolution 234 (1996 August 
2). These observations provided a low-resolution ($\lambda/\Delta\lambda 
\approx$ 90) spectrum between 2.5 to 11.6 $\mu$m.  We processed the
data using the
ISOPHOT Interactive Analysis (PIA) software\footnote{PIA is a joint
development by the ESA astrophysics Division and the ISOPHOT Consortium led
by the Max Planck Institut f\"ur Astronomie (MPIA),
Heidelberg. Contributing ISOPHOT Consortium institutes are DIAS, RAL, AIP,
MPIK and MPIA.}, version 6.0.  Our ISOPHOT-S spectrum of NGC 4151 is 
displayed in Figure \ref{f:PHT-spectrum}.

ISOPHOT-S spectra are ideally suited for studying emission features produced by
`large molecules' such as the Polycyclic
Aromatic Hydrocarbons (PAHs).  Such emission features trace star-formation 
regions and are weak or absent in AGNs (Roche et al. 1991,
Genzel et al. 1998). 
In NGC\,4151 we do not detect any PAH features. We determined upper limits 
to the PAH fluxes at 6.2$\mu$m and 7.7$\mu$m by first subtracting
a continuum set by a linear interpolation between the fluxes at 5.9$\mu$m and 
10.95$\mu$m, and then integrating the spectrum from 6.0 to 6.5 
and 7.3 to 8.2 $\mu$m. 
We derive upper
limits of 2.5e-19 and 3.7e-19 W/cm$^2$ for the 6.2$\mu$m and 7.7$\mu$m
PAH fluxes.  
These limits imply that starburst regions contribute at most
20\% to the production of the low-ionization fine-structure lines in
NGC 4151 (see A98).
Our optical spectral observations were carried out with the KSO Coud\'{e} 
Echelle Spectrometer
during several nights between 6th and 10th of February 1998. We used the UV 
Prism (3600 - 5300 {\AA}) and a slit size of $6.8\arcsec\times15\arcsec$,
corresponding to a spectral resolution of about 10000. The position angle
for the different observations varied between 0 and 40 degrees. We coadded
4 spectra of 1800 seconds exposure time each. Our goal was to obtain a high 
quality large aperture spectral scan of the [OIII] 5007 {\AA} emission line. 
Our result is shown in Figure \ref{f:kso_spe}.


\section{Line Profiles}
\label{s:profiles}

An inspection of Figure \ref{f:SWS-spectrum} shows that all of the detected
fine-structure emission lines are asymmetric, with excess emission clearly 
visible on the blue sides of the lines.  
On average the flux under the red wing of the line profiles is 80\% of 
the blue wing flux. 
In Table 1
we quantify these asymmetries by listing the line center
velocities (C50, e.g. Heckman et al. 1981) defined at 50\% of the peak
intensities, relative to the systemic
velocity of 1000 km s$^{-1}$ (Schulz 1990).  All of the line-center
velocities are negative
indicating a blue-shifted bias in the emission lines.  We find a mean
value $<C50>=-44\pm 20$ km s$^{-1}$ for the mid-IR lines listed in 
section \ref{f:SWS-spectrum}.

Blue asymmetries are well known features of the
optical narrow emission lines in many Seyfert galaxies, including NGC 4151.
Our observations reveal, for the first time, similar asymmetries in the mid-IR 
emission lines.  A key question which we now wish to address is: How similar 
are the mid-IR profiles to the optical emission line profiles?

Many optical emission line observations of NGC 4151 have been presented in 
the literature. These include the ground-based imaging and spectroscopic 
studies by Walker (1968a,b), Ulrich (1973), 
Heckman et al. (1981), Pelat \& Alloin (1982), Schulz (1987, 1990)
and Veilleux (1991a,b,c), and the more recent {\it HST} observations by Evans 
et al. (1993), Boksenberg et al. (1995), Winge et al. (1997) and Hutchings et 
al. (1998).
The biconical structure of the NLR in NGC 4151 was already apparent in 
Walker's and Ulrich's observations. Indeed, the four major [OIII] emission 
complexes identified by Ulrich (see her Figure 4) are discernable (at much 
higher resolution) in the
recent {\it HST} STIS image presented by Hutchings et al. 
Heckman et al. (1981) carried out a long-slit spectroscopic survey of 
[OIII] $\lambda$ 5007\AA
emission in a sample of Seyfert galaxies and found that most of the lines in 
their sample display significant blue asymmetries. In NGC 4151 they
found that C50=-90 km s$^{-1}$ relative 
to the systemic velocity. 
Pelat \& Alloin (1982) carried out high-resolution 
($\sim 15$ km s$^{-1}$) observations
of the [OIII] 5007\AA \,line with it's prominent blue shoulder. 
Similar high-resolution observations 
were carried out by Schulz (1987), who observed 
(narrow) H$\alpha$ and H$\beta$ and 
the [NII] $\lambda\lambda 6548,6583$, [SII] $\lambda 6716,6731$ and [OI] 
$\lambda 6300$ 
lines in addition to the [OIII] doublet.  Schulz 
found an [OIII] profile very similar to that 
measured by Pelat \& Alloin,
and found that 
$<C50>=-37\pm 6$ km s$^{-1}$ averaged
over the lines he observed, in good agreement with the mean value of the 
{\it ISO} mid-IR lines. 
More recently, Veilleux (1991a) carried out 
$\sim 10$ km s$^{-1}$ resolution
and $2.5\arcsec\times 2.5\arcsec$ aperture optical observations of several 
Seyferts. 
For NGC 4151 Veilleux presented profile observations of 
H$\alpha$, H$\beta$ and the
[OI], [OIII], [SII], and [NII] lines, and also the (narrow) HeI $\lambda 
5876$, HeII $\lambda 4686$
lines, as well as the [ArIII] $\lambda 7136$, [FeVII] $\lambda 5721$, and 
[FeVIII] $\lambda 5721$
lines (see his Figure 14).  
All of the lines of NGC\,4151 displayed by Veilleux (1991a) show 
blue asymmetries
similar to those found by Pelat \& Alloin and Schulz,
though there are small differences in detail between the different lines.

Here we compare our mid-IR spectra to the optical line profiles
presented by Veilleux (1991a). 
These optical observations are spatially integrated over the central 
$\sim 3\arcsec$, which is small compared to the SWS apertures. However, 
our large aperture
[O III] 5007{\AA} spectrum (Figure \ref{f:kso_spe}), is almost
identical to the spectrum presented by Veilleux (see also Figures 5a and b).
This demonstrates that our ISO observations are likely sampling the
same regions giving rise to the optical line profiles.

Three further considerations are required in 
this comparison.
First, with the exception of
[ArIII] 8.99 $\mu$m, the mid-IR fine-structure lines are
not emitted by the same ions which emit the optical forbidden lines.  
However, emission lines from neighboring ionization states of the same element 
are available in the combined mid-IR and optical data set.  We therefore chose 
to compare the profiles of the optical/mid-IR line
pairs [OIII]5007{\AA}/[OIV]25.89$\mu$m, [SII]6716{\AA}/[SIII]18.71$\mu$m, 
[SII]6716{\AA}/[SIII] 33.48$\mu$m,
[SII]6716{\AA}/[SIV]$10.51$, and [ArIII]7136{\AA}/[ArIII]8.99$\mu$m.
Second, correlations between the profile shapes and the ionization level,
gas temperature, and transition critical densities could introduce
intrinsic differences between the optical and mid-IR emission lines.
In NGC 4151 such correlations appear to be very weak, and the evidence is
controversial (Veilleux 1991c, Schulz 1990). 
No such correlations are apparent in our ISO observations.
This supports the conclusions of Veilleux (1991b) that the
line shapes do not depend on the ionization potentials
or the critical densities of the transitions.
Third, the SWS spectral resolution is much lower than the resolution of the 
optical measurements.
We therefore convolved the [O III], [S II], and 
[Ar III] optical lines with the appropriate ISO instrumental profiles
at the wavelengths of the [O IV], [S III], [SIV] and [ArIII] fine-structure 
lines.
The instrumental SWS profiles are well approximated by a Gaussian, and 
the variation of profile width with wavelength is known (Valentijn et al. 
1996a). The instrumental profile depends on the source extension, but
because the NLR is compact compared to the SWS aperture we treated the NLR as 
a point source.

The result of our optical-infrared comparison is shown in Figure 
\ref{f:prof_comp}.                                        
It is immediately apparent that the optical and mid-IR profiles are very 
similar. 
Differences between the profiles are no larger than 5-10\% (see 
the residual plots in each panel of Figure \ref{f:prof_comp}). As shown above
the average line center velocities ($<C50>$) are identical within the
uncertainties.
Panels a and b demonstrate the equivalence
of the Veilleux (1991a) data with large aperture data.
The prominent shoulders in the optical profiles are smeared out 
somewhat when smoothing to the lower SWS resolution, but the profiles still 
display pronounced blue asymmetries, as do the mid-IR lines.
In the next section we argue that the remarkable similarities between
the optical and mid-IR profiles point to a specific origin for the
blue asymmetries.

\section{Discussion}
\label{discus}

The simplest models of optical line profile asymmetries in Seyfert galaxies
invoke spherically symmetric outflows or inflows, combined with
extended dust extinction through the NLR or dusty emission line clouds
(Heckman et al. 1981; Dahari \& De Robertis 1988; De Robertis \& Shaw 1990).
In outflow models, the outwardly moving clouds are embedded in an extended 
dusty NLR with a total line-of-sight visual extinction $A_V\sim 1$.
Red-shifted emission is produced by clouds on the far-side and is attenuated
relative to blue-shifted emission on the near-side of the NLR.
In inflow models, the NLR is assumed to be optically thin, but
each of the individual inflowing emission-line clouds are dusty with 
$A_V\sim 1$.
In this picture, the blue-shifted clouds are on the far-side of the NLR and
are visible because their illuminated sides face the observer. Red-shifted 
emission
is produced on the inflowing clouds on the near-side of the NLR, but is 
attenuated by
the dust within the clouds.  However, because $A_V\approx 50 A_{mid-IR}$ 
neither of these models are
compatible with our finding that the optical and mid-IR lines display almost
identical blue asymmetries.  The $A_V\sim 1$ required to produce an asymmetry 
in the optical lines, without completely blocking the red-shifted emission, 
would have negligible distorting effect on the mid-IR lines.
Further, for any value of $A_V$ one would expect the mid-IR lines 
to show less asymmetry than the optical lines.

Alternatively, the blue asymmetries could be produced by very optically thick 
material which is opaque at both optical and mid-IR wavelengths, and which 
fully blocks
some of the more distant NLR emission from view (Whittle 1985; Dahari \& De 
Robertis 1998; Veilleux 1991c; Quintilio \& Viegas 1997).
This possibility is consistent only with outflowing gas, where the 
blue-shifted material is on the near-side of the NLR.  This scenario is 
strongly favored by our observations,
since identical optical and mid-IR line profiles would then be expected. 

Are the blue-shifted NLR clouds in NGC 4151 outflowing? Hutchings et al. (1998)
found that many of the high-velocity blue-shifted [OIII] emission line clouds 
they observed are moving at velocities very close to the velocity of the 
blue-shifted [CIV] absorption features present
in the UV spectrum of NGC 4151 (Weymann et al. 1997). As noted by Hutchings 
et al. (1998) this strongly suggests that the blue-shifted clouds are 
outflowing and are on the near-side of the NLR.

An occulting gas mass may be estimated from the assumption that the NLR 
outflow in NGC 4151 is approximately symmetric, and 
that about 20\% of the red-shifted counterparts to the high velocity 
blue-shifted clouds are blocked from view, in order to reproduce the 
observed line asymmetry.
The blue-shifted high velocity clouds are located within distances 
of the order 1$\arcsec$ to 2$\arcsec$ of
the nucleus on the SW side of the NLR cone, with the highest velocity clouds
closest to the nucleus 
(Hutchings et al. 1998; Ulrich 1973). We approximate the cone as having 
constant surface brightness in the lines.
We hence estimate that, to obscure 20\% of the red-shifted emission, 
a region of (projected) radius $\sim 0.7\arcsec$  has to be blocked by opaque 
material ($(0.7\arcsec/1.5\arcsec)^2 \sim 0.2)$, perhaps in the form of a disk 
or a torus which is perpendicular to the jet axis. 
For this disk or torus to be opaque in the mid-IR its thickness 
must correspond to a hydrogen column density of $\gtrsim 10^{23}$ cm$^{-2}$ 
assuming a Galactic gas-to-dust ratio. The total mass is then 
$\gtrsim 5\times10^6$ M$_\odot$. This is a lower limit, because we did not
consider de-projection of the radius of the obscuring disk and because the 
column density could be much higher.
The lower limit for the occulting gas mass is fully consistent with 
the 2.4x10$^7$ solar masses inferred from 
millimetre CO measurements by Rigopoulou et al. (1997) within a 23$\arcsec$ 
beam (assuming a Galactic CO/H$_2$ conversion factor). Thus the obscuring 
material could be molecular. 

We note that (at least part of) the rotational H$_2$ emission we have observed
could be produced in this obscuring mass. 
The line width (FWHM) of the two detected H$_2$
lines is approximately 230 km/s.
Such a velocity, interpreted as virial flow in a compact region of radius
46 pc, translates into a central mass of 2.7x$10^8$ M$_{\odot}$.
This is an approximate but plausible central mass in view of 
recent results from reverberation mapping (Edelson et al 1996), and also 
compared to the total enclosed
masses within the central 50 pc regions of other galaxies, including
the Milky Way (e.g. Genzel, Hollenbach \& Townes, 1994).
The rotational H$_2$ lines trace 
warm gas at temperatures of $\sim$100 K and higher. For thermal emission
the ratio of the S(1) and S(0) lines is approximately $112{\rm exp}(-505/T)$
where $T$ is the gas temperature (K). Our observed lower limit of 2.4
for this flux ratio implies a minimum temperature of 130 K for the S(1)
emitting gas.  At this temperature the observed S(1) flux corresponds to
a (maximum) warm H$_2$ mass of $3\times 10^7$ M$_\odot$. At temperatures 
closer to 200 K, that appear plausible from observations of other galaxies
(Rigopoulou et al. 1996, Valentijn et al. 1996b), the warm H$_2$ mass
would be $\sim 3\times10^6$ M$_\odot$. This suggests that
the bulk of the obscuring material may be warm.  

It is, of course, possible that the observed anisotropic distribution of the 
line emitting clouds is real, i.e. there could simply be more bright, blue 
shifted clouds close to the nucleus than red shifted ones, giving rise to 
identical optical and mid-IR line asymmetries.
However, it would be difficult to reconcile such an idea with the
preferential blueshift and blueward asymmetry also found in many other Seyferts
(e.g. Heckman et al. 1981, Veilleux 1991b).

\section{Conclusions}
\label{s:Conclusions}

We have obtained ISO SWS and ISOPHOT-S spectra of the NLR of the Seyfert galaxy
NGC\,4151. We detect 17 fine structure emission lines, as well as two lines of
molecular hydrogen (0-0 S(1) and S(5)) and one hydrogen recombination line
(Br$\beta$). In addition we derived upper limits for 4 further fine structure
lines, and H$_2$ S(2) and S(0). The
emission lines span a wide range of ionization potential (8-303 eV), tracing
the ionizing UV radiation in a wavelength range that has barely been accessible
before.

The PAH features are absent within our detection limit. The upper limits
we derive for the features at 6.2 and 7.7 $\mu$m confirm that there is no
significant starburst contribution to the total luminosity of the nuclear
region. 

The emission line profiles are asymmetric (in the sense of a blue excess) 
and blue shifted with respect to the sytemic velocity. We have compared
the ISO line profiles to optical, high resolution profiles. 
The profiles are very similar, reproducing
the same asymmetries and blue shifts. Since the infrared lines are much less
sensitive to extinction, this means that simple dust-plus-radial-motion 
models cannot explain the observed profiles.

Two alternative models remain plausible: a real anisotropic distribution of the
line emitting clouds, and an obscuring screen close to the nucleus.
The first alternative might be a good explanation for a few 
individual objects. In larger samples, however, this effect should average 
out. 
Our prefered model is that of a
geometrically thin but optically thick obscuring screen
of (projected)
sub-arcsecond extension, enclosing a total mass of $\gtrsim 5\times10^6$
M${\odot}$. These dimensions are consistent with HST imaging and
millimetre CO measurements. The obscuring mass may be primarliy
molecular and the source of the H$_2$ rotational emission we
have observed.

Our study demonstrates the value of optical-infrared line profile comparisons
and the potential of applying this method to other Seyfert galaxies observed
with ISO-SWS.

\acknowledgements
We are grateful to Henrik Spoon for providing us with the reduced ISOPHOT-S
spectrum and PAH flux measurements. We also thank Eike G\"unther for
taking the Tautenburg (KSO) spectrum, and Thomas Gehren for providing 
software and assistance with the reduction of it. We thank Mike Eracleous
for dicussion.
This work was supported by DARA under
grants 50-QI-8610-8 and 50-QI-9492-3, and by the German-Israeli Foundation
under grant I-196-137.7/91. 


\clearpage

\figcaption[f1.eps]{The SWS line spectra of NGC\,4151 \label{f:SWS-spectrum}}

\figcaption[f2.eps]{The ISO-SWS aperture position superimposed on the [O III]
$\lambda$5007
contours of the NGC\,4151 nucleus (adapted from Yoshida \& Ohtani 1993).
The interval between the solid contour lines is 1 magnitude. The dotted 
contour is 8.5 magnitudes weaker than the central contour. The rectangles
represent the $14\arcsec\times20\arcsec$ and $20\arcsec\times33\arcsec$ -- i.e.
the smallest and largest -- aperture of SWS. \label{f:aperture}}

\figcaption[f3.eps]{The ISOPHOT-S spectrum of NGC\,4151 (long wavelength part 
only). The dominant pixel at 10.5 $\mu$m is the [S IV] line. There is no 
indication for strong PAH features at 6.2, 7.7, 8.7 or 11.2 $\mu$m, 
and hence no indication for a strong contribution of a starburst component.
\label{f:PHT-spectrum}}

\figcaption[f4.eps]{Large aperture [O III] 5007{\AA} spectrum of NGC\,4151. 
\label{f:kso_spe}}

\figcaption[f5.eps]{Comparison of the ISO SWS line profiles (solid lines) with 
their
optical counterparts convolved with the SWS instrumental profile (dashed). 
The optical lines are taken from: a: The KSO Tautenburg [O III] 5007{\AA} 
spectrum. b: 
[O III] 5007{\AA} of 
V91a. c,d,e,f: [Ar III] and [S II] of V91a. The lower part of each panel shows 
the difference (ISO - optical) of the two profiles. 
+10\% and -10\% margins in the residual plots are shown as horizontal dashed 
lines. 
Units are: velocity (km/s) 
with respect to the peak (x-axis), and normalized flux densities (y-axis).
\label{f:prof_comp}}

\clearpage

\begin{table}
\caption{Observed line fluxes, profile widths, and velocity shifts.}
\label{t:SWS-linefluxes}
\begin{center}
\begin{scriptsize}
\begin{tabular}{lr@{.}lr@{.}lr@{.}lr@{.}lcc}
\hline
Line & \multicolumn{2}{c}{$\lambda_0$} & 
       \multicolumn{2}{c}{$E_{\rm ion}$$^{\rm a}$}&
       \multicolumn{2}{c}{$f_\ell$$^{\rm b}$} & 
       \multicolumn{2}{c}{$\Delta f_\ell$$^{\rm c}$} &
       FWHM$^{\rm d}$ & 
       C50$^{\rm e}$\\ 
     & \multicolumn{2}{c}{$\mu$m} & 
       \multicolumn{2}{c}{eV}&
       \multicolumn{4}{c}{$10^{-13}$ erg s$^{-1}$ cm$^{-2}$} &
       km\,s$^{-1}$ &
       km\,s$^{-1}$\\
\hline
$[$Si\,{\sc ix}$]$   &   2&584 & 303&2 &  0&23& 0&05&  ?  &  ?  \\
Br$\beta$            &   2&625 &  13&6 &  0&47& 0&01&  ?  &  ?  \\
$[$Mg\,{\sc viii}$]$ &   3&028 & 224&9 &  0&62& 0&04& 258 & -22 \\
$[$Si\,{\sc ix}$]$   &   3&935 & 303&2 &  0&41& 0&01& 170 & -29 \\
$[$Mg\,{\sc iv}$]$   &   4&487 &  80&1 &  0&31& 0&02&  ?  &  ?  \\
H$_2$ S(5)           &   6&910 & \multicolumn{2}{c}{---} &  0&85& 0&05& 235 & 
-76 \\
$[$Ne\,{\sc vi}$]$   &   7&652 & 126&2 &  7&83& 0&04& 281 & -25 \\
$[$Ar\,{\sc iii}$]$  &   8&991 &  27&6 &  2&2 & 0&1 & 357 & -79 \\
$[$S\,{\sc iv}$]$    &  10&511 &  34&8 & 11&3 & 0&1 & 354 & -24 \\
$[$Ne\,{\sc ii}$]$   &  12&814 &  21&6 & 11&8 & 0&1 & 318 & -48 \\
$[$Ne\,{\sc v}$]$    &  14&322 &  97&1 &  5&5 & 0&5 & 288 & -44 \\
$[$Ne\,{\sc iii}$]$  &  15&555 &  41&0 & 20&7 & 0&2 & 346 & -51 \\
H$_2$ S(1)           &  17&035 & \multicolumn{2}{c}{---} &  1&20& 0&15& 227 & 
-22 \\
$[$S\,{\sc iii}$]$   &  18&713 &  23&3 &  5&4 & 0&01& 431 & -24 \\
$[$Ne\,{\sc v}$]$    &  24&317 &  97&1 &  5&59& 0&05& 376 & -68 \\
$[$O\,{\sc iv}$]$    &  25&890 &  54&9 & 20&3 & 0&2 & 451 & -83 \\
$[$Fe\,{\sc ii}$]$   &  25&988 &   7&9 &  0&44& 0&02& 447 & -23 \\
$[$S\,{\sc iii}$]$   &  33&480 &  23&3 &  8&1 & 0&5 &  ?  &  ?  \\
$[$Si\,{\sc ii}$]$   &  34&815 &   8&2 & 15&6 & 0&9 & 352 & -42 \\
$[$Ne\,{\sc iii}$]$  &  36&013 &  41&0 &  3&5 & 0&1 &  ?  &  ?  \\
\hline                            
$[$Mg\,{\sc vii}$]$ &    5&50  & 186&5 &$<$1&0 &\multicolumn{2}{l}{---}& --- & 
--- \\
$[$Mg\,{\sc v}$]$    &   5&610 & 109&2 &$<$1&5 &\multicolumn{2}{l}{---}& --- & 
--- \\
H$_2$ S(2)           &  12&279 & \multicolumn{2}{c}{---} &$<$1&2 
&\multicolumn{2}{l}{---}& --- & --- \\
$[$Fe\,{\sc iii}$]$  &  22&925 &  16&2 &$<$0&2 &\multicolumn{2}{l}{---}& --- & 
--- \\
$[$Fe\,{\sc i}$]$    &  24&042 &   0&0 &$<$0&4 &\multicolumn{2}{l}{---}& --- & 
--- \\
H$_2$ S(0)           &  28&219 & \multicolumn{2}{c}{---} &$<$0&5 
&\multicolumn{2}{l}{---}& --- & --- \\
\hline			       	       
PAH                  &   6&2   & \multicolumn{2}{c}{---} 
&$<$25&0&\multicolumn{2}{l}{---}& --- & --- \\
PAH                  &   7&7   & \multicolumn{2}{c}{---} 
&$<$37&0&\multicolumn{2}{l}{---}& --- & --- \\
\hline
\end{tabular}
\begin{itemize}
\item[$^{\rm a}$] Ionization potential of the stage leading to the transition.
\item[$^{\rm b}$] Observed flux.
\item[$^{\rm c}$] Error estimate on observed flux, based only on the
uncertainty in defining the underlying continuum. In addition there is a 
general flux calibration uncertainty of $\approx$20\%.
\item[$^{\rm d}$] Observed width, i.e. without deconvolution of the
instrumental profile. Note that the FWHM of the instrumental profile ranges
between 120 and 300 km/s.
\item[$^{\rm e}$] The unweighted mean velocity of the profile at half maximum
relative to the systemic velocity of 1000 km\,s$^{-1}$ (Schulz
\cite{Schulz87}). Question marks mean that the profile parameters could
not be determined reliably.
\end{itemize}
\end{scriptsize}
\end{center}
\end{table}

\clearpage
\begin{figure}
   \centerline{
   \epsfbox{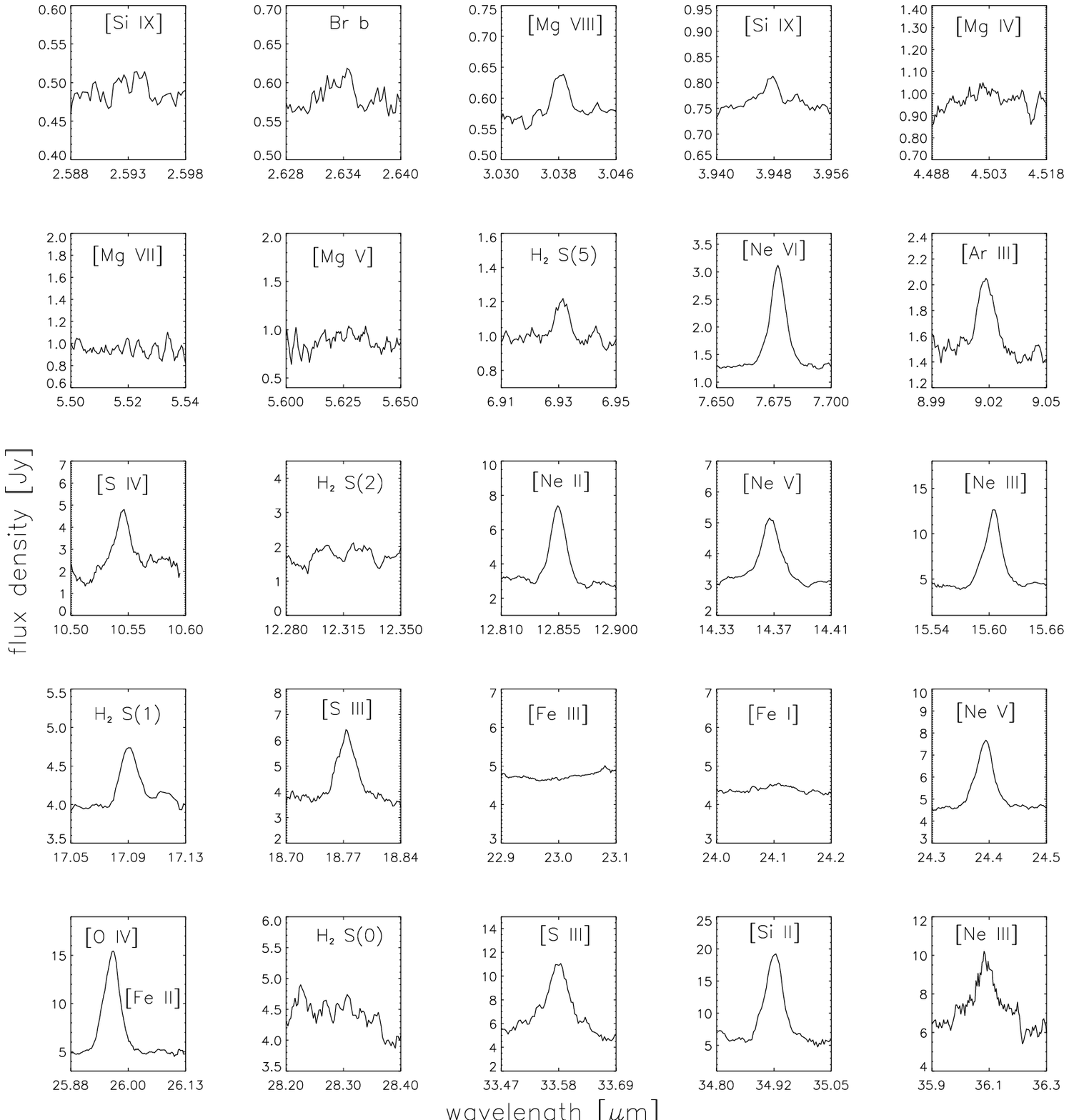} 
   } 
\end{figure} 
%
%
\begin{figure}
   \epsfxsize=275pt
   \epsfbox{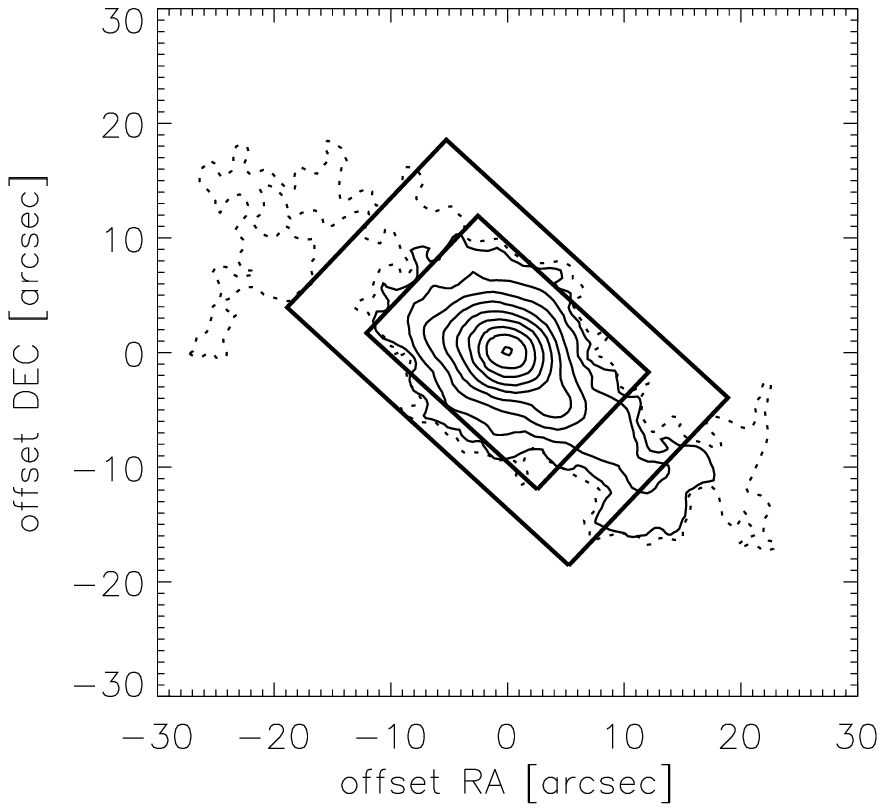}
\end{figure}
%
%
\begin{figure}
   \centerline{
   \epsfbox{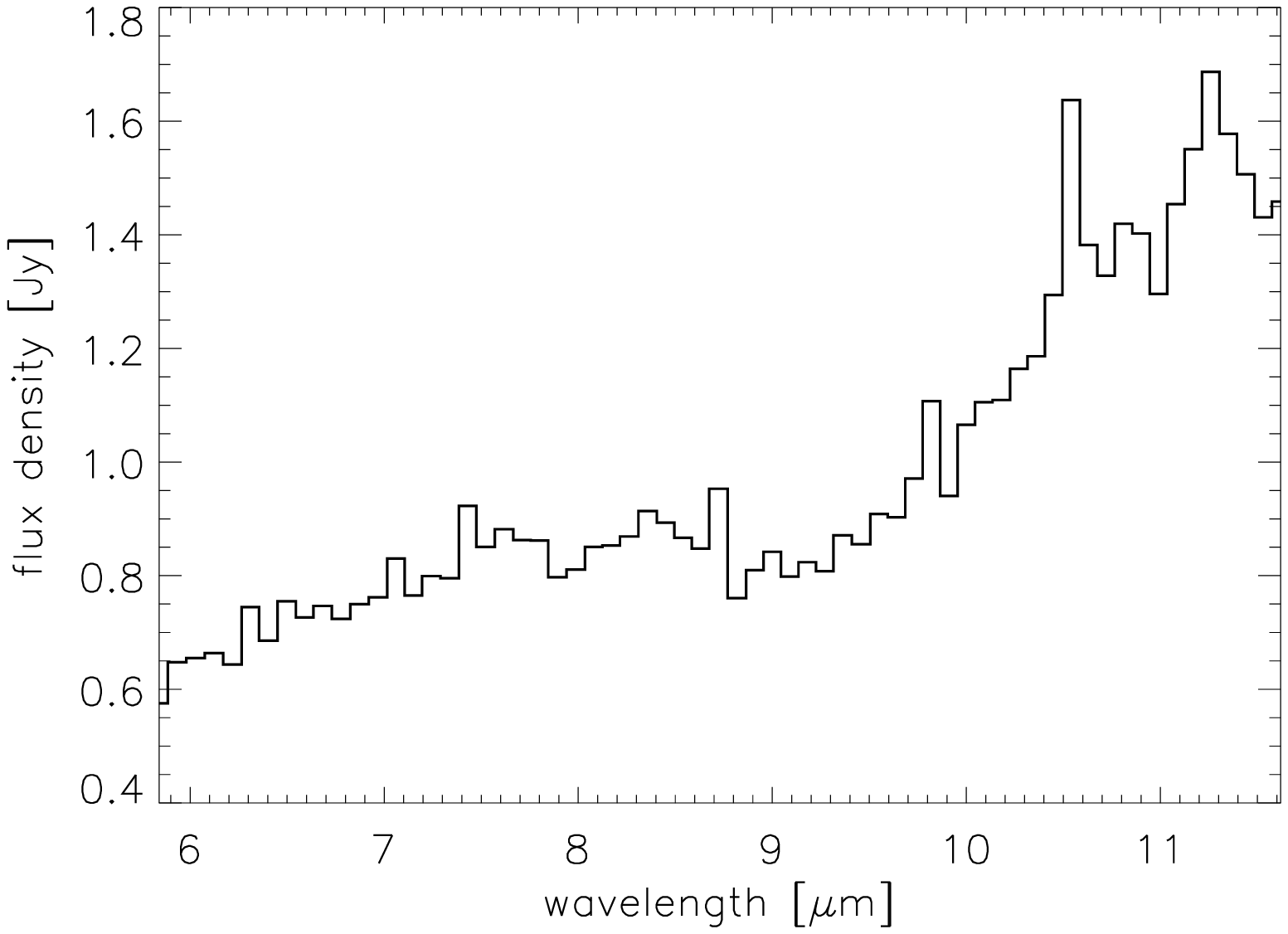}
   } 
\end{figure} 
%
\begin{figure}
   \centerline{
   \epsfbox{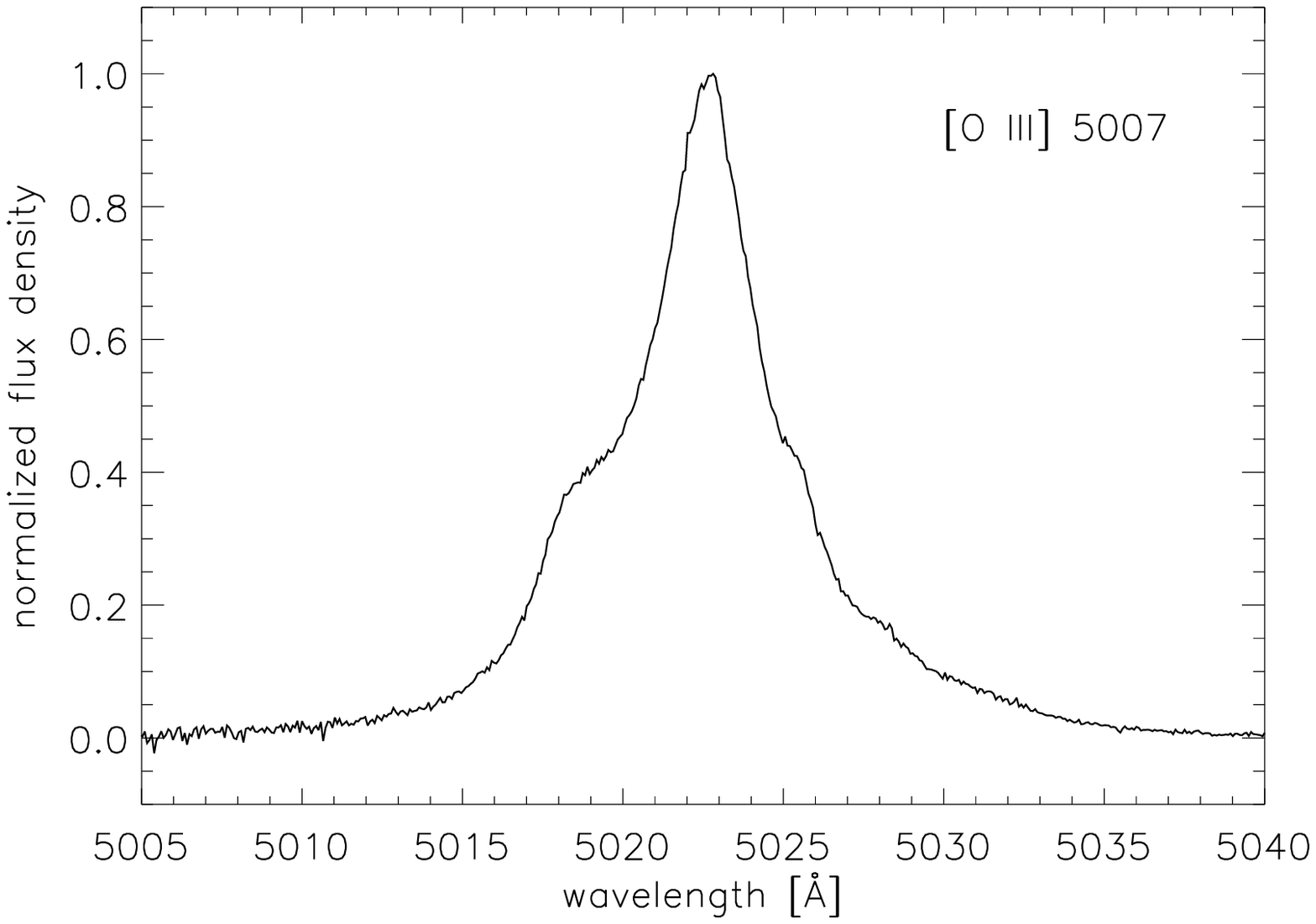}
   } 
\end{figure} 
%
\begin{figure}
   \centerline{
       \epsfbox{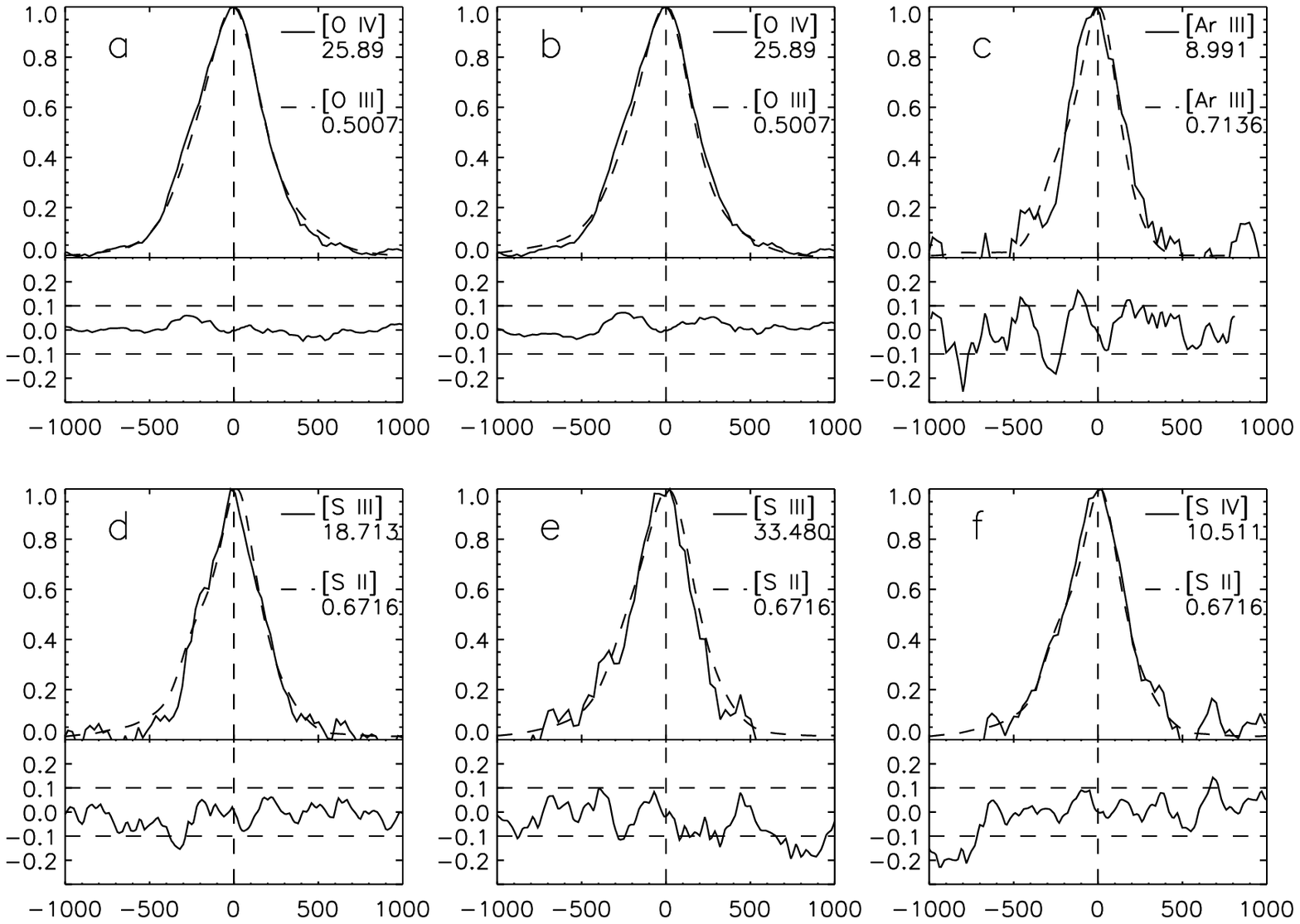}
   }
\end{figure} 
%
\end{document}